\documentclass[aps,prb,twocolumn,groupedaddress,floatfix,showpacs]{revtex4}
\usepackage{subfigure}
\usepackage{color}
\usepackage{graphicx}
\usepackage{amssymb}
\usepackage{amsmath}
\usepackage{natbib}

\begin{document}

\title{Convexity of the self-energy functional in the variational cluster
approximation}
\author{Andriy H. Nevidomskyy}
\email[E-mail: ]{nevidomskyy@cantab.net}
\author{David S\'en\'echal}
\author{A.-M.S. Tremblay}
\affiliation{D\'{e}partment de physique, Universit\'{e} de Sherbrooke, Sherbrooke, Qu\'{e}%
bec, J1K 2R1, Canada}
\date{\today}

\begin{abstract}
In the variational cluster approximation (VCA) (or variational cluster
perturbation theory), widely used to study the Hubbard model, a fundamental
problem that renders variational solutions difficult in practice is its
known lack of convexity at stationary points, i.e. the physical solutions
can be saddle points rather than extrema of the self-energy functional. Here
we suggest two different approaches to construct a convex functional $\Omega
\lbrack \Sigma ]$. In the first approach, one can show analytically that in
the approximation where the irreducible particle-hole vertex depends only on
center of mass coordinates, the functional is convex away from phase
transitions in the corresponding channel. Numerical tests on a tractable
version of that functional show that convexity can be a nuisance when
looking for instabilities both in the pairing and particle-hole channels.
Therefore, an alternative phenomenological functional is proposed. Convexity
is explicitly enforced only with respect to a restricted set of variables,
such as the cluster chemical potential that is known to be otherwise
problematic. Numerical tests show that our functional is convex at the
physical solutions of VCA and allows second-order phase transitions in the
pairing channel as well. This opens the way to the use of more efficient
algorithms to find solutions of the VCA equations.
\end{abstract}

\pacs{71.10.Fd, 
71.15.Dx, 
71.10.-w, 
71.27.+a 
}

\keywords{variational cluster approximation, VCPT, Potthoff's scheme,
convexity}
\maketitle

\parskip=0pt







\section{Introduction}

\label{Sec.intro}

Effective functionals, such as the Landau-Ginzburg free energy functional in
the vicinity of phase transitions, have long been used to study classical
and quantum systems. Typically, the effective functional $F[h]$ is obtained
in terms of a relevant variable $h$, such as the order parameter in the
Landau-Ginzburg theory, or the electron density in the density functional
theory~\cite{DFT}. The optimal value of $h$ is then obtained from the
requirement that the functional be stationary at the solution, $\delta
F/\delta h=0$.

In the formalism of Luttinger Ward~\cite{LW60} and Baym Kadanoff~\cite{BK61}%
, one of the best known functional approaches for correlated electrons, a
functional $\Omega \lbrack G]$ is stationary and equal to the grand
potential at the exact physical value of $G$. In such a scheme, the
functional dependence $\Omega \lbrack G]$ is not known exactly. It can be
approximated perturbatively by summing a subset of the infinite series of
skeleton diagrams that define the Luttinger-Ward functional. To obtain
non-perturbative results on the other hand, one of the most effective
methods is the Dynamical Mean-Field Theory (DMFT) \cite{Georges:1996}.
Chitra and Kotliar \cite{CK01} have shown how this theory can be obtained by
modifying the Kadanoff-Baym functional and making the local approximation on
the stationarity condition. More recently, a general scheme for generating a
wide class of non-perturbative approximations for the Hubbard model \cite%
{Hubbard} from a functional has been proposed by Potthoff~\cite{Potthoff03}.
In this method, known as the self-energy-functional approach (SFA), a new
functional $\Omega \lbrack \Sigma ]$ of the self-energy $\Sigma $ is
constructed, which is stationary at the physical solution. The functional
itself is unknown explicitly, but Potthoff suggested a particular way of
calculating a variational solution $\delta \Omega /\delta \Sigma =0$ with
the help of a reference system, typically a cluster of finite size, which
can be solved exactly. This particular implementation of the self-energy
functional with no bath (contrary to DMFT) goes under the name of the \emph{%
variational cluster approximation} (VCA). DMFT and generalizations thereof,
(such as Cellular Dynamical Mean-Field Theory (CDMFT)) can be obtained as
various special cases of SFA~\cite{{Potthoff03},{Potthoff03b}} corresponding
to different choices of reference systems and/or approximations of the
stationarity condition. (The functionals of Chitra and Kotliar \cite{CK01}
and of Potthoff \cite{{Potthoff03},{Potthoff03b}} are in fact identical, as
shown in Appendix \ref{CK}).

One desirable feature of functional approaches is the variational principle
that guarantees that an approximate grand potential is an upper bound to the
true grand potential. Such a variational principle is missing for the
stationary solutions of both the Baym-Kadanoff functional~\cite%
{{BK61},{Kotliar99}} and Potthoff's self-energy functional~\cite{Potthoff03}
since stationary solutions are known to be saddle points rather than extrema.

A question thus remains open up to present, whether or not it is possible to
construct a functional, be it a functional of the Green function $G$ or of
the self-energy $\Sigma $, such that its stationary solutions would always
be extrema (say, minima) of the functional. If so, this would mean that the
functional is convex at the physical solution, which is a first necessary
step on the way to prove the variational principle. For an
infinite-coordination Bethe lattice, this question was answered positively
by Kotliar~\cite{Kotliar99}, who proved that a functional could be
constructed, such that its extrema occurred at the physical local Green
function of the Hubbard model. However an attempt by Chitra and Kotliar to
find its analogue for a finite-dimensional lattice was only partially
successful~\cite{CK01}.

Another motivation to find convex functionals is a practical one. In VCA,
the functional is definitely not convex for example when the intra-cluster
chemical potential is varied. The physical solutions are then saddle points.
However, most efficient numerical algorithms such as e.g. the conjugate
gradient method, have been designed to find extrema of a functional rather
than saddle points. Although one may attempt to use such algorithms to also
find saddle points (by minimizing the magnitude square of the gradients of
the functional \cite{PotthoffComm}), the unphysical solutions may occur.

In this paper we reexamine the above problem and show that a convex
functional can be found. In particular, we construct a new functional of the
self-energy $\Omega \lbrack \Sigma ]$, such that its stationary solutions
are minima when the irreducible particle-hole vertex depends only on center
of mass coordinates and the system is away from phase transitions in the
corresponding channel. Going beyond this approximation involves complicated
integrals that cannot be treated analytically, however numerical tests on a
tractable version of the functional suggest that it is indeed always convex
at the physical solution. Moreover, we show that such a construction is not
unique and that several functionals can be constructed that differ in the
higher-order terms of the expansion with respect to the self-energy.

Despite the convexity of the proposed functional, its implementation
requires additional approximations and it turns out to be inadequate to
detect a second-order phase transition for both Cooper pairing and
antiferromagnetic instabilities. There, the new functional appears to always
be convex at the \emph{paramagnetic} solution, whereas in the case of
second-order phase transitions, the paramagnetic solution is rightly
expected to be a saddle point, with minima developing instead at a finite
value of the symmetry-breaking order parameter. In other words, the sought
convexity of the functional 'overdoes' its job, imposing too stringent
conditions on the resulting physical solution.

Given that the VCA method has originally been developed to study broken
symmetry phases, the aforementioned feature of the proposed new functional
is especially undesirable. To cure this drawback, we propose a different,
this time phenomenological, approach that ensures convexity at the physical
solution and, moreover, respects the tendency of the system to develop an
instability towards a broken symmetry phase. This finding opens up new
perspectives in the use of the convexity property of the functional, in
particular permitting to apply powerful numerical techniques, such as the
conjugate gradient method, which are only guaranteed to work if the solution
is known to be an extremum (and not a saddle point) of the functional.

The paper is organized as follows. First, in Section~\ref{Sec.review} we
review the variational cluster approximation (VCA) as proposed by Potthoff~%
\cite{Potthoff03}, followed by a general discussion of the stability of the
stationary solution and criteria for convexity in Section~\ref{Sec.stability}%
. We then proceed to derive a new functional of self-energy in Section~\ref%
{Sec.new} with the proof of its convexity given in Appendix~\ref%
{Sec.convexity} and the recipe for incorporating it into the VCA framework
in Appendix~\ref{Sec.implement}. The second part of Section~\ref{Sec.new} is
devoted to numerical tests of the proposed functional on a two-dimensional
Hubbard model, which show that the functional is convex but that it fails to
correctly describe second-order phase transitions into a broken symmetry
phase. Triggered by this negative result, we propose in Section~\ref%
{Sec.altern} an answer to the convexity problem while correcting the
aforementioned failure to describe second-order phase transitions. The
adequacy of this new construction is corroborated by numerical tests. More
technical aspects of the work are detailed in appendices.

\section{Review of the conventional VCA scheme}

\label{Sec.review}

\begin{figure}[tbp]
\par
\begin{center}
{\includegraphics[width=8.6cm]{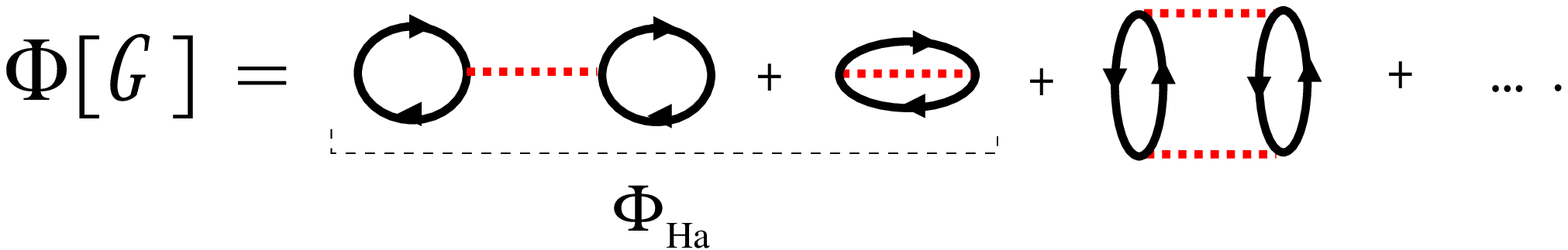}}
\end{center}
\caption{(Color online) Luttinger-Ward functional $\Phi[G]$ constructed as a
sum of renormalised skeleton diagrams with appropriate combinatorial
coefficients (not shown). The Hartree approximation consists in considering
the first two diagrams only of the series. }
\label{Fig.LW}
\end{figure}

Consider, following Potthoff~\cite{Potthoff03,kotliarRMP:2007}, a general
Hamiltonian $H=H_{0}(\mathbf{t})+H_{1}(\mathbf{U})$ with one-particle
hopping parameters $\mathbf{t}$ and two-particle interaction parameters $U$: 
\begin{equation}
H=\sum\limits_{ij}t_{ij}c_{i}^{\dagger }c_{j}+\frac{1}{2}\sum%
\limits_{ijkl}U_{ijkl}c_{i}^{\dagger }c_{j}^{\dagger }c_{l}c_{k}.
\end{equation}%
Here ${i,j,k,l}$ refer to an orthonormal and complete set of one-particle
basis states. The equilibrium thermodynamics and elementary one-particle
excitations of the system for temperature $T$ and chemical potential $\mu $
are fully described by the one-particle Matsubara Green function~\cite%
{Doniach_book} defined by the imaginary-time ordered product $\mathrm{T}$ 
\begin{equation}
G_{\sigma }(\mathbf{r}_{1},\tau _{1};\mathbf{r}_{2},\tau _{2})=-\langle 
\mathrm{T}[c_{\sigma }(\mathbf{r}_{1},\tau _{1})c_{\sigma }^{\dagger }(%
\mathbf{r}_{2},\tau _{2})]\rangle ,
\end{equation}%
where the symbol $\langle \dots \rangle $ denotes thermal average and $%
G_{\sigma }(1,2)\equiv G_{\sigma }(\mathbf{r}_{1},\tau _{1};\mathbf{r}%
_{2},\tau _{2})$ can be seen as $\langle \mathbf{r}_{1},\tau _{1}|\widehat{G}%
_{\sigma }|\mathbf{r}_{2},\tau _{2}\rangle $ or $G_{12},$ a matrix of two
indices, each of which stands for both space and imaginary time.

We start by defining the Luttinger-Ward functional $\Phi \lbrack G]$,
constructed formally as a sum of all closed, irreducible skeleton diagrams
involving fully renormalized (\textquotedblleft dressed\textquotedblright )
Green functions~\cite{LW60}, as illustrated in Fig.~\ref{Fig.LW}. We note
that this functional is universal, that is, it does not depend on the
kinetic part of the Hamiltonian $H_{0}(\mathbf{t})$, but only on the
vertices $\mathbf{U}$ and the Green function itself.

The important property of the Luttinger-Ward functional is that its
functional derivative gives the self-energy of the system: 
\begin{equation}
\frac{\delta \Phi \lbrack G]}{\delta G_{\sigma }\left( 1,2\right) }=\Sigma
_{\sigma }\left( 2,1\right) =\Sigma _{\sigma }[G].  \label{LW}
\end{equation}%
This equation serves as a definition of the self-energy as a functional of $%
G $. We can now introduce the grand potential of the system as the
Baym-Kadanoff functional~\cite{BK61} defined in terms of the fully dressed
Green function as follows: 
\begin{equation}
\Omega _{BK}[G]=\Phi \lbrack G]-\mathrm{Tr}\left( \left(
G_{0}^{-1}-G^{-1}\right) G\right) +\mathrm{Tr}\ln (G).  \label{BK}
\end{equation}%
Inverting Eq. (\ref{LW}) (locally \cite{Potthoff03}) to obtain the Green
function as a functional of $\Sigma $, we can now express the Luttinger-Ward
functional as a functional of self-energy $\Phi \left[ G[\Sigma ]\right] $.
Substituting this into the expression for the Baym--Kadanoff functional (\ref%
{BK}) and using Dyson's equation $G^{-1}=G_{0}^{-1}-\Sigma $, Potthoff
proposed the following functional of the self-energy%
\begin{equation}
\Omega \lbrack \Sigma ]=\Phi \left[ G[\Sigma ]\right] -\mathrm{Tr}\left(
\Sigma G\right) -\mathrm{Tr}\ln (G_{0}^{-1}-\Sigma ).
\end{equation}

\noindent Recognizing that the first two terms on the right hand side of the
Potthoff functional represent the Legendre transform of the Luttinger-Ward
functional 
\begin{equation}
F[\Sigma ]\equiv \Phi \left[ G[\Sigma ]\right] -\mathrm{Tr}\left( \Sigma
G\right) ,  \label{FSigma}
\end{equation}%
one can now rewrite the expression for the grand potential as follows: 
\begin{equation}
\Omega \lbrack \Sigma ]=F[\Sigma ]-\mathrm{Tr}\ln \left( G_{0}^{-1}-\Sigma
\right) .  \label{GPSigma}
\end{equation}%
Using Eqs. (\ref{LW}) and (\ref{FSigma}), it is easy to show that the
following equation holds: 
\begin{equation}
\frac{\delta F[\Sigma ]}{\delta \Sigma _{\sigma }\left( 1,2\right) }%
=-G_{\sigma }\left( 2,1\right) =-G_{21}[\Sigma ].  \label{dF/dSigma}
\end{equation}%
It can be viewed as the definition of the Green function in terms of the
self-energy $\Sigma $. Using this, we immediately arrive at the conclusion
that the variational derivative of the grand potential in Eq. (\ref{GPSigma}%
) vanishes at the true physical solution: 
\begin{equation}
\left. \frac{\delta \Omega \lbrack \Sigma ]}{\delta \Sigma }\right\vert
_{sol}=-G+\left. (G_{0}^{-1}-\Sigma )^{-1}\right\vert _{sol}=0
\label{variation}
\end{equation}%
by virtue of the Dyson equation. At the physical solution, $\Omega \lbrack
\Sigma ]=\Omega \lbrack G]$ is equal to the true grand potential. Therefore
solving the problem amounts to finding such a function $\Sigma $ that
satisfies the above stationarity condition.

Since the Luttinger-Ward functional $\Phi \lbrack G]$ is a universal
functional of the interaction $\mathbf{U}$ and of the stationary value of $G$%
, its Legendre transform $F[\Sigma ]$ is a universal functional of $\mathbf{U%
}$ and $\Sigma $. In other words the value of this functional should not
depend on one-body operators such as the hopping matrix $\mathbf{t}$. This
last fact is the crucial point that allows one to define VCA.

One proceeds as follows. In the case of the Hubbard model where the
interaction is local, one can modify the hopping matrix elements to
subdivide the infinite cluster into disjoint identical clusters. One can
then compute the grand potential $\Omega ^{\prime }$ and the self-energy $%
\Sigma ^{\prime }$ of that problem exactly (for example by means of exact
diagonalization). The one-body part of the cluster Hamiltonian can contain a
different hopping matrix, along with site energies, chemical potential and
Weiss fields, all of which are used as variational parameters. Since one can
write the grand potential for the cluster problem as 
\begin{equation}
\Omega ^{\prime }[\Sigma ^{\prime }]=F^{\prime }[\Sigma ^{\prime }]-\mathrm{%
Tr}\ln \left( G_{0}^{\prime -1}-\Sigma ^{\prime }\right) ,
\end{equation}%
the universality of the functional $F[\Sigma ]$ allows one to find its 
\textit{exact} value for the solution of the cluster problem:%
\begin{equation}
F[\Sigma ^{\prime }]\equiv F^{\prime }[\Sigma ^{\prime }]=\Omega ^{\prime }+%
\mathrm{Tr}\ln \left( G_{0}^{\prime -1}-\Sigma ^{\prime }\right) .
\label{FS'}
\end{equation}%
Using this last result,~the functional $\Omega \lbrack \Sigma ]$ in Eq.~(\ref%
{GPSigma}) can be evaluated exactly when $\Sigma \rightarrow \Sigma ^{\prime
}$ 
\begin{equation}
\Omega \lbrack \Sigma ^{\prime }]=\left( \Omega ^{\prime }[\Sigma ^{\prime
}]+\mathrm{Tr}\ln (G_{0}^{\prime -1}-\Sigma ^{\prime })\right) -\mathrm{Tr}%
\ln (G_{0}^{-1}-\Sigma ^{\prime }).  \label{Omega}
\end{equation}

Since they are the solutions of the cluster problems, the self-energies $%
\Sigma ^{\prime }$ can be varied through the one-body terms of the clusters
(which we define to also contain Weiss fields for various order parameters).
These are collectively represented by the matrix $t_{ij}^{\prime }$.
Clearly, the self-energies obtained in this way will span only a small
subspace of an infinite-dimensional space of all possible variations, namely
only those that can be represented as the physical self-energies of a
cluster $\Sigma ^{\prime }(t^{\prime })$ parametrized by the matrix $%
t_{ij}^{\prime }$. The corresponding stationary solution is obtained by
searching for values of $t_{ij}^{\prime }$ such that 
\begin{equation}
\frac{\mathrm{d}\Omega }{\mathrm{d}t_{i,j}^{\prime }}\equiv \frac{\delta
\Omega }{\delta \Sigma }\cdot \frac{\mathrm{d}\Sigma }{\mathrm{d}%
t_{i,j}^{\prime }}=0.  \label{d_cluster}
\end{equation}%
The set of equations (\ref{Omega})--(\ref{d_cluster}) forms the essence of
the VCA quantum cluster method.

\section{Stability of the stationary solution}

\label{Sec.stability}

Let us consider fluctuations around the stationary solution of an arbitrary
self-energy functional: 
\begin{eqnarray}
\delta \Omega &=&\Omega \lbrack \Sigma +\delta \Sigma ]-\Omega \lbrack
\Sigma ]  \label{fluctuations} \\
&=&\int \!\!\!\int \delta \Sigma _{\sigma }(1^{\prime },1)\frac{\delta
^{2}\Omega }{\delta \Sigma _{\sigma }(1^{\prime },1)\delta \Sigma _{\sigma
^{\prime }}(2^{\prime },2)}\delta \Sigma _{\sigma ^{\prime }}(2^{\prime },2).
\notag
\end{eqnarray}%
The spin indices have been written explicitly and must be summed over. From
now on, we will not write them explicitly to have a lighter notation.

For a stationary point to be numerically stable, it must be a minimum. A
maximum will also do since it suffices to change the sign. Correspondingly,
the functional derivative in Eq.~(\ref{fluctuations}) must be negative or
positive-definite, respectively. It is easy to verify that for Potthoff's
self-energy functional, Eq.~(\ref{GPSigma}), the second functional
derivative is given by 
\begin{equation}
\frac{\delta ^{2}\Omega }{\delta \Sigma _{11^{\prime }}\delta \Sigma
_{22^{\prime }}}=\Gamma _{11^{\prime };22^{\prime }}+G_{1^{\prime
}2}G_{2^{\prime }1},  \label{d2_Omega}
\end{equation}%
where $\Gamma _{11^{\prime };22^{\prime }}$ stands for the second functional
derivative of the universal functional $F[\Sigma ]$: 
\begin{equation}
\Gamma _{11^{\prime };22^{\prime }}\equiv \frac{\delta ^{2}F}{\delta \Sigma
_{11^{\prime }}\delta \Sigma _{22^{\prime }}}  \label{Gamma}
\end{equation}%
and is thus a tensor of the fourth rank.

It follows directly from the expression for the second functional derivative
Eq.~(\ref{d2_Omega}) that the stationary solution is a saddle point.
Following the arguments of Ref.\ \cite{CK01}, this is simplest to illustrate
in the absence of two-body interactions in the Hamiltonian, when $\Sigma =0$
and $F[\Sigma ]=0$. Then the first term in Eq.~(\ref{d2_Omega}) vanishes
while in Matsubara-Fourier space the last term leads to 
\begin{equation}
\frac{\delta ^{2}\Omega }{\delta \Sigma \delta \Sigma }=G\left(
k_{1}^{\prime }\right) G\left( k_{2}^{\prime }\right) \delta \left(
k_{1}^{\prime }-k_{2}\right) \delta \left( k_{2}^{\prime }-k_{1}\right) .
\end{equation}%
In the sector of zero total momentum and total energy $\left( k_{1}^{\prime
}=-k_{2}^{\prime }\right) ,$ this quantity decouples in $2\times 2$ blocks
with zero diagonal elements and equal off-diagonal components equal to $%
(\omega _{n}^{2}+\varepsilon _{k_{1}}^{2})^{-1}.$ The eigenvalues are thus
both positive and negative, which corresponds to a saddle point. The effect
of interactions, $\Gamma _{11^{\prime };22^{\prime }}$, cannot cure this
problem for all wave vectors and frequencies.

\begin{figure}[tbp]
\par
\begin{center}
{\includegraphics[width=8.9cm]{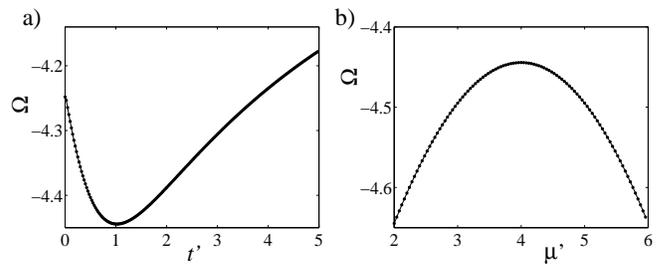}}
\end{center}
\par
\vspace{-4mm}
\caption{Functional $\Omega $ for the 2D Hubbard model ($U/t=8$) calculated
from the exact-diagonalization solution of a 2x2 cluster using Eq.~(  
\protect\ref{Omega}), plotted as a function of (a) cluster nearest-neighbour
hopping parameter $t^{\prime }$ and (b) cluster chemical potential $\protect%
\mu ^{\prime }$. The variational solution of the problem in the space of
cluster parameters $(t^{\prime }=1,\protect\mu ^{\prime }=4)$ is a saddle
point instead of an extremum. }
\label{Fig.old_Omega}
\end{figure}

In the case of VCA, the relevant question concerns stability with respect to
variations in the cluster parameters. It has been pointed out empirically in
the original VCA proposal~\cite{Potthoff03}, that variations with respect to
site energies or chemical potentials lead to saddle points. This is
illustrated in Fig.~\ref{Fig.old_Omega} where the grand potential obtained
from a $2\times 2$ cluster solution is plotted as a function of the
intra-cluster nearest-neighbor hopping $t^{\prime }$ (Fig.~\ref%
{Fig.old_Omega}a) and of the cluster chemical potential $\mu ^{\prime }$
(Fig.~\ref{Fig.old_Omega}b). Clearly, the stationary solution is a minimum
in the former case and a maximum in the latter. Yet, varying the cluster
chemical potential $\mu ^{\prime }$ has been shown~\cite{vary-mu-Aichhorn06}
essential in the VCA scheme for obtaining thermodynamic consistency, which
requires that the number $\langle n\rangle $ of electrons in the system has
the same value when calculated from the two independent relations: 
\begin{equation}
\langle n\rangle =-\frac{\partial \Omega }{\partial \mu }=-\frac{1}{\pi }%
\int\limits_{-\infty }^{\infty }\mathrm{d}\omega \,f(\omega )\,\mathrm{Im}%
\left[ \mathrm{Tr}G(\omega +i0^{+})\right] ,
\end{equation}%
where $f(\omega )=[\exp (\omega /T)+1]^{-1}$ is the Fermi function.
Therefore, it is preferable to let $\mu ^{\prime }$ vary and somehow deal
with the fact that the grand potential is known to be non-convex in this
case.

The fact that the solution is a saddle point rather than an extremum has
consequences for practical implementations, since there seem to exist no
robust numerical search algorithms for a saddle point, whereas many such
algorithms have been developed for extrema~\cite{NumRec} searches. It would
therefore be desirable from this point of view, as well as for the reasons
outlined in Section~\ref{Sec.intro}, to find such a functional $\Omega
\lbrack \Sigma ]$ whose stationary solutions would be guaranteed to be
extrema. In the next Section, we shall demonstrate that such a functional
can indeed be formally constructed, and its convexity can be proved
rigorously given certain approximations.

\section{Convex functionals $\boldsymbol{\Omega[\Sigma]}$}

\label{Sec.new}

In this section, we first derive a general expression for a convex
functional. The drawback of this approach is that in practical
implementations, convexity is preserved for order parameters as well. This
makes it an undesirable feature in the presence of phase transitions. In the
next section, we will restrict ourselves to preserving convexity as a
function of the cluster chemical potential, which in practice is the main
problem to be solved.

\subsection{Analytical results for convex functionals}

\label{Sec.construct}

The main idea of the current approach is to add a term $\Delta \Omega
\lbrack \Sigma ]$ to the grand potential that is quadratic in $\Sigma $ and
that does not alter the stationary solution of the original potential $%
\Omega \lbrack \Sigma ]$, but renders it a positive-definite functional of
the self-energy. We first introduce an auxiliary functional $\mathrm{f}%
[\Sigma ]$ that is defined by 
\begin{eqnarray}
\mathrm{f}[\Sigma ] &=&\mathrm{f}\left( 1,2\right) \equiv \frac{\delta
\Omega \lbrack \Sigma ]}{\delta \Sigma \left( 2,1\right) }  \label{f} \\
&=&\frac{\delta F[\Sigma ]}{\delta \Sigma \left( 2,1\right) }+\left(
G_{0}^{-1}-\Sigma \right) ^{-1}\left( 1,2\right)  \notag
\end{eqnarray}%
and that, by virtue of the stationarity condition Eq.(\ref{variation}),
vanishes identically at the stationary solution of $\Omega \lbrack \Sigma ]$.

It turns out that the simplest term $\Delta\Omega[\Sigma]$ that can be added
to the grand potential is of the form 
\begin{eqnarray}
\Delta\Omega_1 & =& -\frac{1}{2} \mathrm{Tr}\left(\left[ (G_0^{-1}-\Sigma)%
\mathrm{f} \right]^2 \right)  \notag \\
& = & -\frac{1}{2} \mathrm{Tr} \left( 1+(G_0^{-1}-\Sigma)\frac{\delta F}{%
\delta \Sigma} \right)^2.  \label{dOmega1}
\end{eqnarray}

\noindent Indeed, $\Delta \Omega _{1}$ and its functional derivative both
vanish at the stationary solutions of the grand potential, as follows from
the definition of the functional $\mathrm{f}[\Sigma ]$ and, therefore, the
new functional 
\begin{equation}
\Omega _{1}[\Sigma ]=\Omega \lbrack \Sigma ]+\Delta \Omega _{1}[\Sigma ]
\label{new_Omega}
\end{equation}%
yields the same stationary solution as the original one. The proof of the
convexity of this new functional in the special case where the irreducible
particle-hole vertex is local is given in the Appendix~\ref{Sec.convexity}.

We note in passing that our choice of $\Delta \Omega \lbrack \Sigma ]$ is
not unique. To second order in the quantity $\mathrm{f}[\Sigma ]$, the
correction $\Delta \Omega _{1}$ given by Eq.~(\ref{dOmega1}) is the same as,
for example, the following one 
\begin{eqnarray}
\lefteqn{\Delta \Omega _{2}=\mathrm{Tr}\ln \left( 1-(G_{0}^{-1}-\Sigma )%
\mathrm{f}\right) +\mathrm{Tr}\left( (G_{0}^{-1}-\Sigma )\mathrm{f}\right) }
\label{dOmega2} \\
&=&\mathrm{Tr}\ln \left( (G_{0}^{-1}-\Sigma )\frac{\delta F}{\delta \Sigma }%
\right) +\mathrm{Tr}\left( 1+(G_{0}^{-1}-\Sigma )\frac{\delta F}{\delta
\Sigma }\right) ,  \notag
\end{eqnarray}%
meaning that it leaves the stationary point of the original functional
unchanged and leads to exactly the same expression for the second functional
derivative as Eq.~(\ref{dOmega1}). Note also that despite being similar in
spirit to the work of Chitra and Kotliar~\cite{CK01}, the above derivation
is significantly different, as demonstrated in Appendix~\ref{Sec.CK} where
the connection between the Chitra--Kotliar result and the Potthoff
functional is exposed.

While implementing Eqs.~(\ref{dOmega1}, \ref{dOmega2}) in practice, a
problem arises since Eq.~(\ref{f}) for the auxiliary functional $\mathrm{f}%
[\Sigma ]$ contains a functional derivative $\delta F[\Sigma ]/\delta \Sigma 
$ whose value is not readily available in practical calculations. Therefore,
an approximation must be made in order to implement the new functional into
the VCA scheme. One such elegant approximation is given in Appendix~\ref%
{Sec.implement}, and the conclusions of the numerical verification of its
convexity is the subject of the following subsection.

\subsection{Drawback of this type of functionals}

\label{Sec.numerics}

\begin{figure}[tbp]
\par
\begin{center}
{\includegraphics[width=8.9cm]{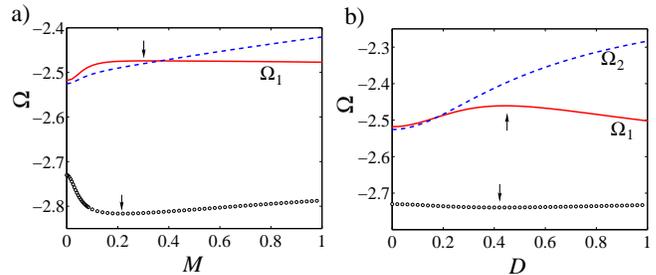}}
\end{center}
\par
\vspace{-4mm}
\caption{(Color online) Dependence of the original Potthoff's functional $%
\Omega $ (open circles), functional $\Omega _{1}$ (solid red line) and $%
\Omega _{2}$ (broken blue line) on the symmetry-breaking Weiss fields
conjugate to (a) antiferromagnetic staggered magnetization, (b) d-wave
superconductivity. Results are for the Hubbard model ($U/t=4$). Arrows
denote positions of minima of $\Omega $ and maxima of $\Omega _{1}$. }
\label{Fig.msupra}
\end{figure}

While numerical tests show convincingly (see Appendix~\ref{Sec.implement}
for details) that the functionals proposed in Eqs.~(\ref{dOmega1})~(\ref%
{Omega2}) are indeed convex with respect to all cluster variational
parameters, including the cluster chemical potential $\mu ^{\prime }$, there
is a major drawback. The numerical tests show one undesirable property,
namely its failure to correctly predict second-order phase transitions, at
least for the approximation to $\mathrm{f}[\Sigma ]$ proposed in Appendix %
\ref{Sec.implement}.

Consider Fig.~\ref{Fig.msupra}a that shows the dependence of $\Omega $ and
of the the new functionals given by Eqs.~(\ref{Omega1}),~(\ref{dOmega2}) on
the magnitude of the antiferromagnetic Weiss field $M$. Unlike the original
functional, which has a minimum at a finite value of $M$, both new
functionals have a single minimum at $M=0$, thereby favouring the
paramagnetic (PM) solution. The behavior as a function of the
superconducting d-wave Weiss field, $D$, follows the same pattern, as shown
in Fig.~\ref{Fig.msupra}b. Only the original functional $\Omega $ exhibits a
minimum at non-zero value of $D$, whereas both functionals proposed in this
work have no other minima except at $D=0$. Clearly, such behavior of the new
functionals is unphysical since the existence of antiferromagnetism in the
Hubbard model, for example around $U=4$ at half-filling, is solidly
established.

As discussed in Appendix~\ref{Sec.convexity}, we would have expected that at
least in the particle-hole channel, instabilities (such as
antiferromagnetism) could have been detected by the present functional.
However, we had to approximate $\delta F[\Sigma ]/\delta \Sigma $ to do the
calculation. In some sense, the requirement of convexity here `overdoes' its
job by rendering the paramagnetic solution always a minimum and hence
ignoring a possible instability towards a broken-symmetry phase!

\section{Curing the problem of convexity arising from a restricted set of
variables}

\label{Sec.altern}

In practical implementations of VCA, the cluster chemical potential leads to
a first-order saddle point even in paramagnetic states. There exist several
numerical techniques that can deal with this type of problems \cite%
{quasi-Newton}. In the context of VCA we show a practical way to transform
the saddle point coming from variation of the cluster chemical potential $%
\mu ^{\prime }$ into a minimum while preserving the `right' of the system to
develop a second-order phase transition. It seems to work well even in cases
where the analytically-obtained approximations given in Eqs.~(\ref{Omega1}),
(\ref{Omega2}) fail.

Let us recall that the general form of the convex correction developed in
Section~\ref{Sec.new} was 
\begin{equation}
\Delta \Omega \sim \mathrm{Tr}\left( \frac{\delta \Omega }{\delta \Sigma }%
\right) ^{2}  \label{dOm1}
\end{equation}%
Typically, we know from experience about the existence of a certain
variational parameter $h$ (here, the cluster chemical potential $\mu
^{\prime }$) such that the $\Omega $ lacks convexity at the stationary
solution $h=h_{0}$. Let us modify the correction in Eq.~(\ref{dOm1}) as
follows: 
\begin{equation}
\Delta \Omega \sim \mathrm{Tr}\left( \frac{\delta \Omega }{\delta \Sigma }%
\cdot \frac{\partial \Sigma }{\partial h}\right) ^{2}=\mathrm{Tr}\left( 
\frac{\partial \Omega }{\partial h}\right) ^{2},  \label{dOm2}
\end{equation}%
where we have effectively substituted the unknown functional derivative $%
\delta \Omega /\delta \Sigma $ by a much simpler derivative with respect to
the variational parameter $h$, which can be easily calculated numerically.

\begin{figure}[tbp]
\par
\begin{center}
{\includegraphics[width=8.9cm]{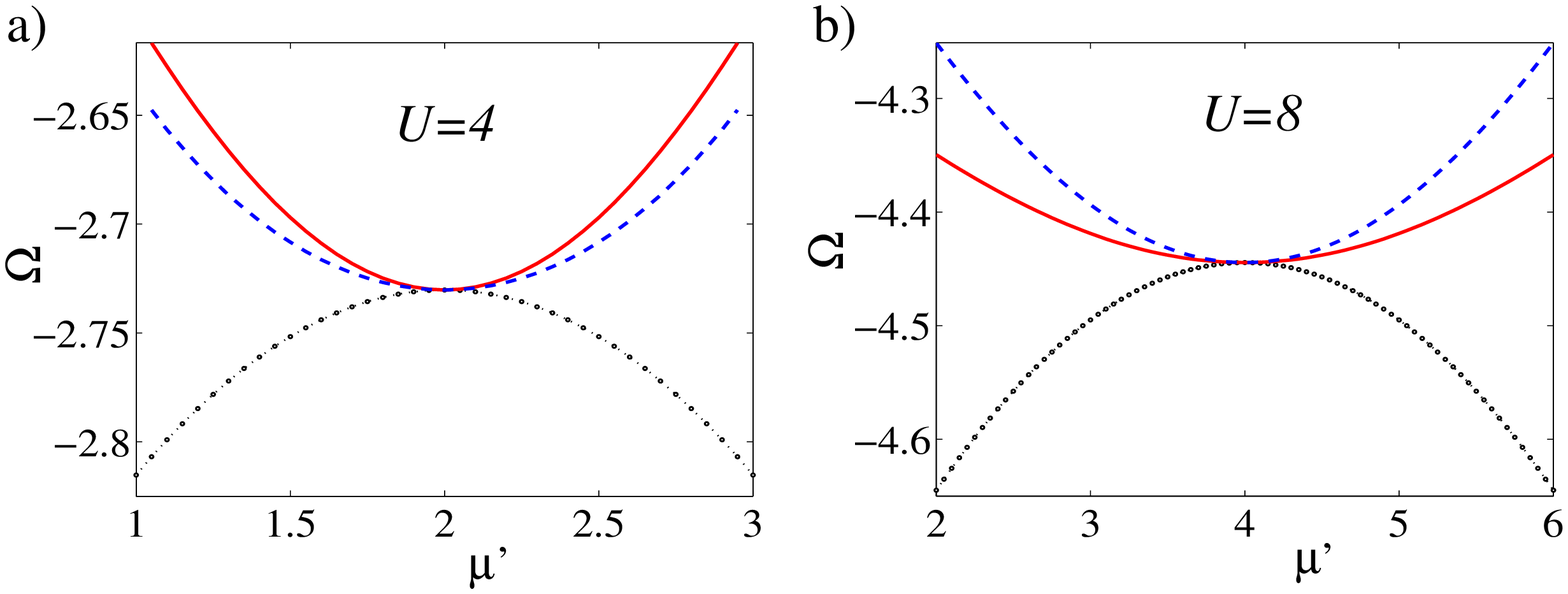}}
\end{center}
\par
\vspace{-4mm}
\caption{(Color online) Dependence on the cluster chemical potential $%
\protect\mu ^{\prime }$ of Potthoff's $\Omega $ (black dots), and the
functional $\Omega _{\protect\lambda }$ in Eq.~(\protect\ref{Omega-new}) for
two choices of the coefficient $\protect\lambda $: the constant $\protect%
\lambda =15$ (solid red line) and the variable $\protect\lambda =2\protect%
\lambda _{0}(\protect\mu ^{\prime })$ (dashed blue line). Results for the
half-filled Hubbard model with (a) $U/t=4$ and (b) $U/t=8$ are shown. }
\label{Fig.mu-new}
\end{figure}

We thus postulate the following functional 
\begin{equation}
\Omega _{\lambda }=\Omega +\frac{\lambda }{2}\left( \frac{\partial \Omega }{%
\partial h}\right) ^{2},  \label{Omega-new}
\end{equation}%
where $\lambda $ is some empirical coefficient that should be chosen such
that the resulting potential $\Omega _{\lambda }$ is a convex function of $h$%
. By construction, the additional term vanishes at the stationary solution $%
h=h_{0}$ where $\partial \Omega /\partial h=0$, and the value of the new
potential coincides with the old one $\Omega (h_{0})$. For the second
derivative at $h_{0}$ we have: 
\begin{equation}
\left. \frac{\partial ^{2}\Omega _{\lambda }}{\partial h^{2}}\right\vert
_{h=h_{0}}=\frac{\partial ^{2}\Omega }{\partial h^{2}}\left( 1+\lambda \frac{%
\partial ^{2}\Omega }{\partial h^{2}}\right) .
\end{equation}%
If the original functional is already convex, \text{$\partial ^{2}\Omega
/\partial h^{2}>0$}, the new functional will be convex too. If however $%
\Omega $ has a maximum and not a minimum at $h_{0}$, the coefficient $%
\lambda $ has to satisfy the following inequality to ensure that the new
functional is convex: 
\begin{equation}
\lambda >\lambda _{0}(h)\equiv \left\vert \left( \frac{\partial ^{2}\Omega }{%
\partial h^{2}}\right) ^{-1}\right\vert .  \label{lambda}
\end{equation}%
In particular, note that choosing $\lambda =2\lambda _{0}(h)$ will result in
the new functional being convex everywhere and having the same absolute
value of curvature as the original one: $\partial ^{2}\Omega _{\lambda
}/\partial h^{2}=-\partial ^{2}\Omega /\partial h^{2}$. Chosen in such a
way, the behavior of the newly constructed functional $\Omega _{\lambda }$
as a function of the cluster chemical potential $h\equiv \mu ^{\prime }$ is
plotted in Fig.~\ref{Fig.mu-new}.

As we see, $\Omega _{\lambda }$ can clearly be made convex with a suitable
choice of the coefficient $\lambda $. The lowest allowed value $\lambda _{0}$
that yields a convex functional is itself a function of parameters of the
model, such as Hubbard $U$, as is clear from comparison of Fig.~\ref%
{Fig.mu-new}a) and b). In particular, we find $\lambda _{0}=5.8$ for $U=4t$
and $\lambda _{0}=9.9$ for $U=8t$.

Note that by taking derivatives with respect to cluster parameters in Eq.~(%
\ref{dOm2}) instead of the more general Eq. (\ref{dOm1}), we have
substituted a strong requirement of convexity with respect to all variations
in $\Sigma $ with a much weaker one, which only requires that the functional
be convex with respect to a particular parameter (or set of parameters) $h$.
Let us then see what effect this has on the dependence of $\Omega _{\lambda
} $ on a symmetry-breaking parameter, such as the superconducting Weiss
field $D$. The first derivative with respect to $D$ reads: 
\begin{equation}
\frac{\partial \Omega _{\lambda }}{\partial D}=\frac{\partial \Omega }{%
\partial D}+\lambda \frac{\partial ^{2}\Omega }{\partial h\partial D}\frac{%
\partial \Omega }{\partial h}  \label{deriv.D}
\end{equation}%
All the stationary points of $\Omega $ (where derivatives with respect to
both $\mu ^{\prime }$ and $D$ vanish) are also stationary points of the new
functional $\Omega _{\lambda }$. That functional can also have additional
stationary points that are not stationary points of $\Omega ,$ but with the
procedure suggested below to choose $\lambda ,$ we found this not to be an
issue.

\begin{figure}[tbp]
\par
\begin{center}
{\includegraphics[width=8.6cm]{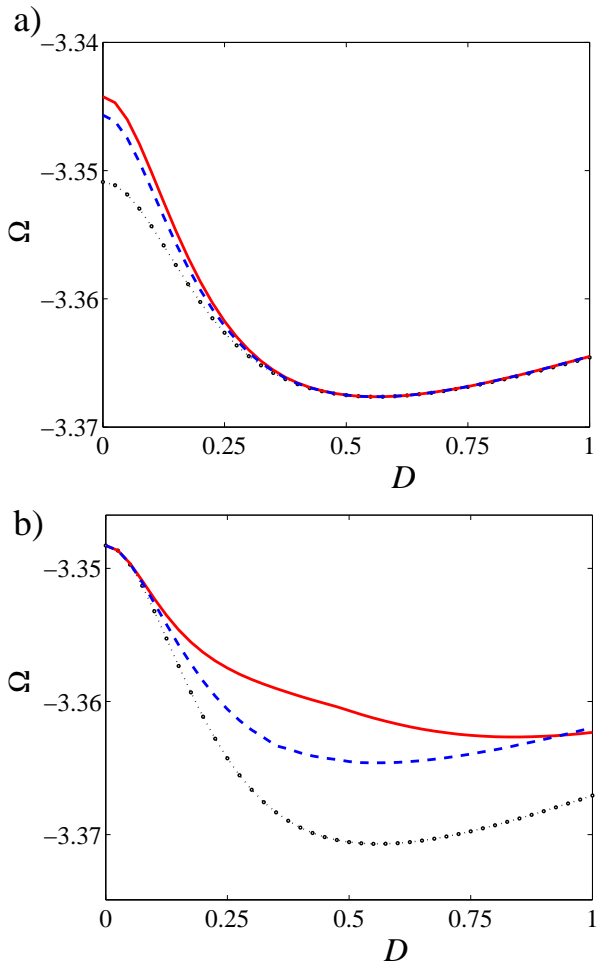}}
\end{center}
\par
\vspace{-5mm}
\caption{(Color online) Dependence on the symmetry-breaking superconducting
Weiss field $D$ of Potthoff's grand potential (black dots), and of the
functional in Eq.~(\protect\ref{Omega-new}) constructed for two choices of
the coefficient $\protect\lambda$: the constant $\protect\lambda=15$ (solid
red curve) and the variable $\protect\lambda=2\protect\lambda_0(\protect\mu%
^{\prime})$ (dashed blue curve). The calculations are for the electron-doped
Hubbard model (2.2\% doping) with $U=4t$ and the next-nearest neighbour
hopping $t^{\prime}=0.3t$. The cluster chemical potential $\protect\mu%
^{\prime}$ was fixed at the value corresponding to (a) the true SC solution, 
$D=0.56$, and (b) the PM solution, $D=0$. }
\label{Fig.supra-new}
\end{figure}

In numerical calculations, we searched for minima of $\Omega _{\lambda }$
and found that, indeed, one recovers as minima the same solutions $(\mu
_{0}^{\prime },D_{0})$ that were saddle points of $\Omega $. To illustrate
how one can choose $\lambda $ in practice, we consider several cases. First,
in Fig.~\ref{Fig.supra-new}a $\mu ^{\prime }$ is fixed at the known
superconducting solution $\mu _{0}^{\prime }$ and $\lambda $ is taken either
as constant or defined by the procedure in Eq.~(\ref{lambda}) that
guarantees convexity. All curves have their minimum at the same value $%
D=D_{0}$ as expected. To gain more insight into the convergence process, we
can also check whether the solution for $D$ is stable in cases where the
value of the cluster chemical potential $\mu ^{\prime }$ is slightly off the
true solution $\mu _{0}^{\prime }$. Such a situation is illustrated in Fig.~%
\ref{Fig.supra-new}b where the chemical potential $\mu ^{\prime }$ is fixed
at the PM value $\mu _{\text{pm}}^{\prime }=2.45$ instead of the true SC
solution $\mu _{0}^{\prime }=2.28$. As expected from Eq.~(\ref{Omega-new}),
both the old and the new functionals coincide at the paramagnetic solution $%
D=0$ since by construction, $\partial \Omega /\partial \mu ^{\prime }=0$
there. The situation is different however away from $D=0$. Fixing $\lambda =%
\mathrm{const}$ (solid line in Fig.~\ref{Fig.supra-new}b) yields a minimum
at an incorrect value of $D=0.84$ instead of $D_{0}=0.56$. By contrast, a
variable coefficient $\lambda =2\lambda _{0}(\mu ^{\prime })$, where $%
\lambda _{0}$ is determined at each point from Eq.~(\ref{lambda}), gives a
minimum of $\Omega _{\lambda}(D)$ (dashed line in Fig.~\ref{Fig.supra-new}b)
that is already very close to the true solution. Varying $\mu ^{\prime }$ as
well eventually leads to the correct solution in all cases, as mentioned
above. An empirical case-by-case analysis has shown that the choice $\lambda
=2\lambda _{0}(\mu ^{\prime })$ gives consistently reliable results and is
preferred over fixing $\lambda $ to a constant value. Besides, such a choice
avoids the ambiguity in the value of $\lambda $ and ensures that the
functional $\Omega _{\lambda }[\mu ^{\prime }]$ is always a convex one,
according to the inequality~in Eq.~(\ref{lambda}). 

The knowledge of the derivatives $\partial \Omega /\partial h$ and $\partial
^{2}\Omega /\partial h^{2}$ is required in order to implement the convex
correction~(\ref{Omega-new}) with $\lambda =2\lambda _{0}(h)$. Numerically,
this requires the knowledge of the functional $\Omega $ at three points. For
this reason, the computational cost of evaluating $\Omega $ at a given point
in parameter space is three times that of the original functional $\Omega $.
However, the minima of the proposed functional may now be searched more
efficiently using powerful numerical methods designed for extrema search,
such as the conjugate gradient method~\cite{NumRec}.

\section{Conclusions}

\label{Sec.conclusions}

It is known that physical solutions obtained with VCA are in general saddle
points rather than extrema of the functional $\Omega \lbrack \Sigma ]$. This
is so in particular in the important practical case where the cluster
chemical potential is varied. Saddle points are notoriously more difficult
to find numerically than extrema. We have thus constructed a new functional
that we proved to be a convex functional of the self-energy at the
stationary solution, at least in the case where the irreducible
particle-hole vertex depends only on the centre of mass coordinates
(Appendix~\ref{Sec.convexity}). It turns out, however, that implementing the
proposed functional in practice is far from simple, since it involves, as
can be seen in Eq.~(\ref{dOmega1}), an unknown functional derivative $\delta
F/\delta \Sigma $ of the Legendre transform $F[\Sigma ]$ of the
Luttinger-Ward functional. An approximation therefore had to be made to
express this functional derivative in terms of cluster-defined quantities,
as detailed in Appendix~\ref{Sec.implement}. The corresponding numerical
results in Sec.~\ref{Sec.numerics} show that the new functional is indeed
convex even in the general non-perturbative case. However attractive this
may be, the proposed functional changes qualitatively the behavior of $%
\Omega $ with respect to symmetry-breaking order parameters, such as
magnetism or superconductivity, rendering instead the paramagnetic solution
stable.

To cure this problem we proposed a functional Eq.~(\ref{Omega-new}) and a
(non-unique) recipe to find the coefficient $\lambda $ that guaranties
convexity with respect to cluster chemical potential $\mu ^{\prime }$ only.
This approach removes the saddle point normally associated with $\mu
^{\prime },$ has the same physical solutions as the original problem and
leaves unchanged the minimum or maximum character of the functional with
respect to variations of symmetry-breaking order parameters, such as
magnetism or superconductivity.

Despite the non-uniqueness of the proposed convex functional and the
admittedly phenomenological basis of its derivation, we argue that this is
an important result. In particular, our findings open a way to the use of
powerful numerical algorithms for solving for minima, such as e.g. conjugate
gradients, that only work provided the functional is convex at the physical
solution. We consider the use of such efficient algorithms as highly
desirable for the VCA approach. The question of the existence of a
functional that would give a bound for the true grand potential and would
thereby implement the variational principle, is left open.

We thank G. Kotliar for useful comments on the manuscript. Computations were
performed on the Elix and the Ms RQCHP clusters. The present work was
supported by FQRNT (Qu\'{e}bec), NSERC (Canada), CFI (Canada), CIAR, the
Tier I Canada Research Chair Program (A.-M.S.T.). 

\appendix

\section{Convexity of the analytical functional}

\label{Sec.convexity}

We aim to prove the convexity of the new proposed functional $\Omega
_{i}[\Sigma ]$ given by either of Eqs.~(\ref{dOmega1}) $\left( i=1\right) $
or (\ref{dOmega2}) $\left( i=2\right) $. At the true solution where $\delta
\Omega _{i}/\delta \Sigma =\mathrm{f}[\Sigma ]=0$, we have that%
\begin{equation}
\left. \frac{\delta ^{2}\Omega _{i}}{\delta \Sigma _{11^{\prime }}\delta
\Sigma _{22^{\prime }}}\right\vert _{\mathrm{f=0}}=\left. \frac{\delta
^{2}\Omega }{\delta \Sigma _{11^{\prime }}\delta \Sigma _{22^{\prime }}}-%
\mathrm{Tr}\left[ G^{-1}\frac{\delta \mathrm{f}}{\delta \Sigma _{11^{\prime
}}}G^{-1}\frac{\delta \mathrm{f}}{\delta \Sigma _{22^{\prime }}}\right]
\right\vert _{\mathrm{f=0}}
\end{equation}%
with $G^{-1}=(G_{0}^{-1}-\Sigma )$ and matrix multiplication implied for the
indices not explicitly written. We then use the definitions%
\begin{eqnarray}
\frac{\delta \mathrm{f}_{12}}{\delta \Sigma _{34}} &=&\frac{\delta ^{2}F}{%
\delta \Sigma _{21}\delta \Sigma _{34}}+\frac{\delta G_{12}}{\delta \Sigma
_{34}}  \notag \\
&=&\Gamma _{21;34}+G_{13}G_{42}.
\end{eqnarray}%
where $\Gamma _{21;34}$ has the symmetry $\Gamma _{21;34}=\Gamma _{34;21},$
and obtain, 
\begin{eqnarray}
\lefteqn{\frac{\delta ^{2}\Omega _{i}}{\delta \Sigma _{11^{\prime }}\delta
\Sigma _{22^{\prime }}}=-\Gamma _{11^\prime ;22^\prime }}
\label{d2_Omega_new} \\
&&-(G_{0}^{-1}-\Sigma )_{33^{\prime }}\cdot \Gamma _{43^{\prime };11^{\prime
}}\cdot (G_{0}^{-1}-\Sigma )_{44^{\prime }}\cdot \Gamma _{34^{\prime
};22^{\prime }}  \notag
\end{eqnarray}%
with the summation implied over the repeated indices. The last term may be
rewritten in the following short-hand matrix notation:

\begin{figure}[tbp]
\par
\begin{center}
{\includegraphics[width=8cm]{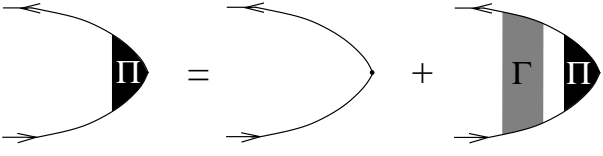}}
\end{center}
\caption{Bethe-Salpeter equation for the particle-hole vertex $\Pi $ in the
language of Feynman diagrams.}
\label{Fig.BS}
\end{figure}

\begin{widetext}%

\begin{equation}
(G_{0}^{-1}-\Sigma )_{33^{\prime }}\cdot \Gamma _{43^{\prime };11^{\prime
}}\cdot (G_{0}^{-1}-\Sigma )_{44^{\prime }}\cdot \Gamma _{34^{\prime
};22^{\prime }}=\mathrm{Tr}\left[ G^{-1}\Gamma _{;11^{\prime
}}^{T}G^{-1}\Gamma _{;22^{\prime }}^{T}\right] .
\end{equation}%

\end{widetext}where the transpose applies only to the indices involved in
the matrix mutiplication.

Irreducible vertices are usually easier to approximate. So it is useful to
work with the second functional derivative of the Luttinger-Ward functional $%
\Phi \lbrack G]$ with respect to the single-particle Green function: 
\begin{equation}
\Gamma ^{\mathrm{ph}}{}_{11^{\prime };22^{\prime }}\equiv \frac{\delta
^{2}\Phi \lbrack G]}{\delta G_{1^{\prime }1}\,\delta G_{2^{\prime }2}}=\frac{%
\delta \Sigma _{11^{\prime }}}{\delta G_{2^{\prime },2}},
\end{equation}%
where we have used the property Eq.~(\ref{LW}) of the Luttinger-Ward
functional. The superscript in the notation $\Gamma ^{\mathrm{ph}}$
indicates that this symmetric matrix $\left( \Gamma _{12;34}^{\mathrm{ph}%
}=\Gamma _{34;12}^{\mathrm{ph}}\right) $ is the \emph{irreducible
particle-hole vertex}, i.e.: 
\begin{equation}
\Pi _{12;34}^{-1}=\delta _{13}\delta _{24}-G_{22^{\prime }}\Gamma
_{2^{\prime }3^{\prime };34}^{\mathrm{ph}}G_{3^{\prime }1},
\end{equation}%
where $\Pi $ is the full (reducible) particle-hole vertex. In the language
of Feynman diagrams, the latter can be cast as shown in Fig.~\ref{Fig.BS}.

It is required on physical grounds that the reducible particle-hole vertex $%
\Pi $ should be positive-definite at zero frequency for stability in the
particle-hole channel and hence it follows that its inverse obeys the same
property. If there is an instability that affects the particle-hole channel
directly or indirectly, it should still be visible in $\Pi _{i}^{-1}.$

Using the identity $\delta F/\delta \Sigma =-G$ and the definition of $%
\Gamma $ given by Eq.~(\ref{Gamma}), we obtain (with summation over repeated
indices implied) that 
\begin{eqnarray}
\Gamma _{12;1^{\prime }2^{\prime }}\Gamma _{1^{\prime }2^{\prime };34}^{%
\mathrm{ph}} &=&\frac{\delta (-G)_{21}}{\delta \Sigma _{1^{\prime }2^{\prime
}}}\cdot \frac{\delta \Sigma _{1^{\prime }2^{\prime }}}{\delta G_{43}} \\
&=&-\delta _{2,4}\delta _{1,3}=-I,
\end{eqnarray}%
which again has the structure of matrix multiplication if the pairs of
indices on either side of the semi-colon are flattened (combined as one).
Using this result we obtain%
\begin{eqnarray}
\mathrm{Tr}(G^{-1}\Gamma _{;11^{\prime }}^{T}G^{-1}\Gamma _{;22^{\prime
}}^{T})\mathrm{Tr}(G\Gamma _{22^{\prime };}^{\mathrm{ph}}G\Gamma
_{33^{\prime };}^{\mathrm{ph}}) &=& \\
\delta _{1,3}\delta _{1^{\prime },3^{\prime }} &=&I.  \notag
\end{eqnarray}%
Using this last identity and Eq.~(\ref{d2_Omega_new}) for $\delta ^{2}\Omega
_{i}/\delta \Sigma ^{2}$, we find:%
\begin{widetext}
\begin{equation}
-\frac{\delta ^{2}\Omega }{\delta \Sigma _{11^{\prime }}\delta \Sigma
_{22^{\prime }}}\mathrm{Tr}(G\Gamma _{22^{\prime };}^{\mathrm{ph}}G\Gamma
_{33^{\prime };}^{\mathrm{ph}})=\delta _{1,3}\delta _{1^{\prime },3^{\prime
}}-G_{14}\Gamma _{44^{\prime };33^{\prime }}^{\mathrm{ph}}G_{4^{\prime
}1^{\prime }}.  \label{identity}
\end{equation}%
\end{widetext}%
As has been pointed out above, the right-hand side of this equation is
positive-definite at zero frequency, following from the requirement that the
kernel of the particle-hole vertex equation be positive.

Let us now concentrate on $L_{22^{\prime };33^{\prime }}[G]=\mathrm{Tr}%
(G\Gamma _{22^{\prime };}^{\mathrm{ph}}G\Gamma _{33^{\prime };}^{\mathrm{ph}%
})$ appearing on the left hand side of the last equation~(\ref{identity}).
We assume that the irreducible particle-hole vertex is local (like in the
Random Phase Approximation), i.e.

\begin{equation}
\Gamma _{11^{\prime };22^{\prime }}^{\mathrm{ph}}=\delta _{11^{\prime
}}\delta _{22^{\prime }}\widetilde{\Gamma }_{12}^{\mathrm{ph}}.
\end{equation}%
With this approximation, we can write for the left-hand side of the
stability equation (\ref{identity}):%
\begin{widetext}
\begin{eqnarray}
-\frac{\delta ^{2}\Omega }{\delta \Sigma _{11}\delta \Sigma _{22}}%
G_{6^{\prime }5^{\prime }}\Gamma _{22;5^{\prime }5^{\prime }}^{\mathrm{ph}%
}G_{5^{\prime }6^{\prime }}\Gamma _{6^{\prime }6^{\prime };33}^{\mathrm{ph}}
&=&-\frac{\delta ^{2}\Omega }{\delta \Sigma _{11}\delta \Sigma _{22}}%
\widetilde{\Gamma }_{25^{\prime }}^{\mathrm{ph}}G_{6^{\prime }5^{\prime
}}G_{5^{\prime }6^{\prime }}\widetilde{\Gamma }_{6^{\prime }3}^{\mathrm{ph}}
\\
&=&-\frac{\delta ^{2}\Omega }{\delta \Sigma _{11}\delta \Sigma _{22}}%
\widetilde{L}_{23}
\end{eqnarray}%
Taking the Fourier-Matsubara transform, we find%
\begin{eqnarray}
\widetilde{L}\left( \mathbf{Q,}i\omega _{\nu }\right) &=&\widetilde{\Gamma }%
^{\mathrm{ph}}\left( \mathbf{Q,}i\omega _{\nu }\right) \left[ \int_{\mathbf{k%
}}\sum_{i\omega _{n}}G\left( \mathbf{k},i\omega _{n}\right) G\left( \mathbf{%
k+Q},i\omega _{n}+i\omega _{\nu }\right) \right] \widetilde{\Gamma }^{%
\mathrm{ph}}\left( \mathbf{Q,}i\omega _{\nu }\right) \\
&=&-\widetilde{\Gamma }^{\mathrm{ph}}\left( \mathbf{Q,}i\omega _{\nu
}\right) \chi \left( \mathbf{Q},i\omega _{\nu }\right) \widetilde{\Gamma }^{%
\mathrm{ph}}\left( \mathbf{Q,}i\omega _{\nu }\right)  \label{Valeur de L}
\end{eqnarray}%
\end{widetext}%
where the dressed Lindhard function $\chi (\mathbf{Q},i\omega _{\nu })$ (no
vertex correction) denotes the value of the integral (defined with the minus
sign) in the last equation. 
Using the spectral representation one can show that $\chi \left( \mathbf{Q}%
,i\omega _{\nu }\right) $ for a system at equilibrium is always positive for
all Matsubara frequencies. Also, $\widetilde{\Gamma }^{\mathrm{ph}}\left( 
\mathbf{Q,}i\omega _{\nu }\right) $ is real, as follows from its spectral
properties, so $\left( \widetilde{\Gamma }^{\mathrm{ph}}\left( \mathbf{Q,}%
i\omega _{\nu }\right) \right) ^{2}$ is positive. It is thus clear that the
quantity $\widetilde{L}(\mathbf{Q},i\omega _{\nu })$ must be
negative-definite, with possible exception of phase transitions that, in
this simple approximation, could appear in the particle-hole channel.
Comparing this with the identity given by Eq.~(\ref{identity}): 
\begin{equation}
-\frac{\delta ^{2}\Omega _{i}}{\delta \Sigma \delta \Sigma }\cdot
L[G]=1-G\Gamma ^{\mathrm{ph}}G,
\end{equation}%
\begin{widetext}%
\begin{equation}
-\frac{\delta ^{2}\Omega _{i}}{\delta \Sigma \delta \Sigma }\left( \mathbf{Q,%
}i\omega _{\nu }\right) \cdot \widetilde{L}\left( \mathbf{Q,}i\omega _{\nu
}\right) =1+\chi \left( \mathbf{Q},i\omega _{\nu }\right) \widetilde{\Gamma }%
^{\mathrm{ph}}\left( \mathbf{Q},i\omega _{\nu }\right)
\end{equation}%
\end{widetext}%
and remembering that its right hand side is positive-definite, we arrive at
the conclusion that, at least in the above approximation for the
particle-hole vertex, the object 
$\delta ^{2}\Omega _{i}/\delta \Sigma ^{2}$ must also be positive-definite
at zero frequency. This in turn means that our proposed grand potential $%
\Omega _{i}[\Sigma ]$, defined by Eq.~(\ref{new_Omega}) and either
expressions (\ref{dOmega1}) or (\ref{dOmega2}), is a convex functional of
the self-energy within this approximation.

It should be noted that being a convex functional of the self-energy is not
necessarily the same as being a convex functional of a given cluster
parameter $h^{\prime }$. Indeed, differentiating again the first derivative $%
\mathrm{d}\Omega /\mathrm{d}h^{\prime }$ given by Eq.~(\ref{d_cluster}), one
obtains 
\begin{equation}
\frac{\mathrm{d}^{2}\Omega _{i}}{\mathrm{d}h^{\prime }{}^{2}}%
=\int\limits_{1}\!\!\int\limits_{2}\frac{\delta ^{2}\Omega _{i}}{\delta
\Sigma (1)\delta \Sigma (2)}\frac{\mathrm{d}\Sigma (2)}{\mathrm{d}h^{\prime }%
}\frac{\mathrm{d}\Sigma (1)}{\mathrm{d}h^{\prime }}+\int\limits_{1}\frac{%
\delta \Omega _{i}}{\delta \Sigma (1)}\frac{\mathrm{d}^{2}\Sigma (1)}{%
\mathrm{d}h^{\prime }{}^{2}}.  \label{d2_cluster}
\end{equation}%
Note that in the above, we have assumed translational invariance, and hence
self-energy can be written as a function of one momentum and frequency
variable only. Since the functional derivative in the first term of Eq.~(\ref%
{d2_cluster}) is almost certainly positive-definite at the stationary
solution, as suggested above, the first term is guaranteed to be positive
too. The situation with the second term is more complicated. Naively, it may
seem that this term vanishes since, at the stationary solution, $\delta
\Omega /\delta \Sigma $ is zero by definition. This would be true if the
variational space of the cluster parameter $h^{\prime }$ was sufficient to
describe the complete variational space of the functional $\Omega \lbrack
\Sigma (h^{\prime })]$. Unfortunately, and this is the main approximation
entering the VCA method, this may not always be true. In fact, one may only
vary $\delta \Sigma (h^{\prime })$ in a subspace that can be parametrized by
the cluster parameter $h^{\prime }$, which is a small part of the whole
infinite-dimensional space of variation $\delta \Sigma $. Therefore, the
solution obtained by requiring that the derivative $\partial \Omega
/\partial h^{\prime }$ vanishes, is not necessarily the same as the true
solution where $\delta \Omega /\delta \Sigma =0$. Although we expect this
contribution to be very small, it is difficult to judge the convexity of the
given grand potential with respect to some cluster parameter, even if the
potential is known to be a convex functional of $\Sigma $. Since varying $%
h^{\prime }$ is the best one can achieve within the VCA approach we stress
that numerical tests are always desirable to verify the convexity of a given
functional $\Omega \lbrack \Sigma (h^{\prime })]$.

\section{Implementation of the convex functional in VCA}

\label{Sec.implement}

In Section~\ref{Sec.new} we proposed a correction $\Delta \Omega _{1,2}$ to
the original grand potential $\Omega \lbrack \Sigma ]$, such that it is
rendered a convex functional of the self-energy. The issue that we address
here is how to implement this correction in the framework of the VCA quantum
cluster method.

The key element entering the expressions (\ref{dOmega1}) and (\ref{dOmega2})
for $\Delta \Omega $ is the auxiliary functional $\mathrm{f}[\Sigma ]$
defined by Eq.~(\ref{f}) as a functional derivative $\mathrm{f}[\Sigma
]\equiv \delta \Omega /\delta \Sigma $. Unfortunately it is not possible to
evaluate this functional derivative even numerically since we have no direct
way of varying the self-energy $\Sigma $ (however we develop a variant of
this idea further in Section~\ref{Sec.altern}). The unknown element in $%
\mathrm{f}[\Sigma ]$ Eq.~(\ref{f}) is the functional derivative $\delta
F[\Sigma ]/\delta \Sigma $, where $F[\Sigma ]$ is a universal functional of
the self-energy, as explained in section~\ref{Sec.review}. This functional
derivative can be evaluated at the \emph{ground state solution of the cluster%
} by virtue of Eq.~(\ref{dF/dSigma}): 
\begin{equation}
\frac{\delta F[\Sigma ]}{\delta \Sigma }\approx \left. \frac{\delta F[\Sigma
^{\prime }]}{\delta \Sigma ^{\prime }}\right\vert _{sol^{\prime
}}=-G^{\prime }\,,  \label{dF/dSigma'}
\end{equation}%
where, as before, we denote the quantities belonging to the cluster by
prime. Now, that both terms on the right hand side of Eq.~(\ref{f}) are
known, we can write: 
\begin{equation}
\mathrm{f}[\Sigma ]\approxeq -G^{\prime }+\frac{1}{G_{0}^{-1}-\Sigma }\equiv
-G^{\prime }+G\,.  \label{f'}
\end{equation}%
We stress that this last equation is approximate. This is because the
self-energy $\Sigma ^{\prime }$ of the cluster solution, at which the
derivative in (\ref{dF/dSigma'}) is evaluated, is not, generally speaking,
equal to the \emph{true lattice} self-energy. Another way to see this is to
note that, by construction, $\mathrm{f}[\Sigma ]$ must vanish identically at
the true stationary solution \emph{of the lattice}, whereas it is clear that
the right hand side of Eq.~(\ref{f'}) can only vanish if the two Green
functions are equal to each other at the ground state \emph{of the cluster
tiling}. Rigorously speaking, this can only be the case in the limit of
infinitely large cluster, where $G^{\prime }\rightarrow G$.

Using this approximation Eq.(\ref{f'}), we can now rewrite Eq.~(\ref{dOmega1}%
) for $\Delta \Omega _{1}$ as 
\begin{equation}
\Delta \Omega _{1}[\Sigma ]=-\frac{1}{2}\mathrm{Tr}\left[ 1-(G_{0}^{-1}-%
\Sigma )G^{\prime })\right] ^{2}\,,  \label{dOmega1'}
\end{equation}%
so that, by virtue of Eq.~(\ref{Omega}), the new convex functional $\Omega
_{1}=\Omega +\Delta \Omega _{1}$ finally becomes: 
\begin{eqnarray}
\Omega _{1}[\Sigma ]\approxeq \Omega ^{\prime } &-&\mathrm{Tr}\ln \left[
(G_{0}^{-1}-\Sigma )G^{\prime }\right]  \notag \\
&-&\frac{1}{2}\mathrm{Tr}\left[ 1-(G_{0}^{-1}-\Sigma )G^{\prime })\right]
^{2}\,.  \label{Omega1}
\end{eqnarray}

\noindent Similarly, for the correction $\Delta \Omega _{2}$ expressed by
Eq.~(\ref{dOmega2}), we obtain an alternative expression for the new
functional 
\begin{equation}
\Omega _{2}[\Sigma ]\approxeq \Omega ^{\prime }+\mathrm{Tr}\left[
1-(G_{0}^{-1}-\Sigma )G^{\prime }\right] \,.  \label{Omega2}
\end{equation}

\begin{figure}[tbp]
\par
\begin{center}
{\includegraphics[width=8.9cm]{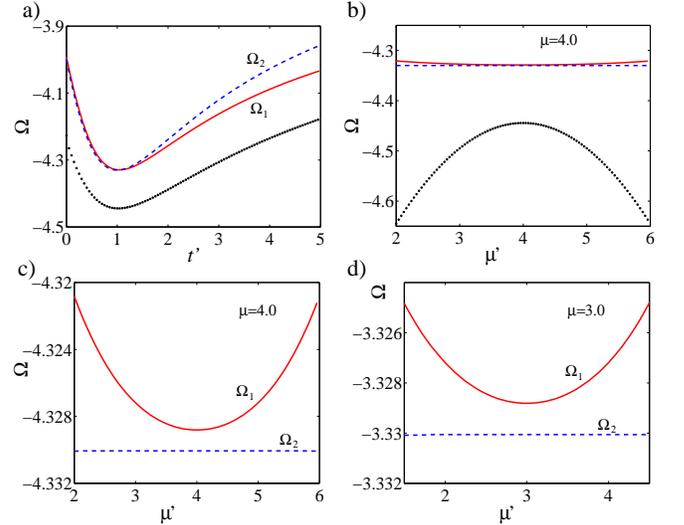}}
\end{center}
\par
\vspace{-3mm}
\caption{(Color online) Comparisons of three different functionals for the
2D Hubbard model ($U/t_{\mathrm{lat}}=8$): the original functional from Eq.~(%
\protect\ref{Omega}) (black dots), and the two functionals proposed in this
work, given respectively by Eq.~(\protect\ref{Omega1}) (solid red curve) and
Eq.~(\protect\ref{Omega2}) (dashed blue curve). The panels show (a)
dependence of all three functionals on the nearest-neighbour hopping
parameter $t^{\prime }$ of the cluster, (b) same for the dependence on the
chemical potential $\protect\mu ^{\prime }$ of the cluster. Comparison
between the two functionals proposed in this work as functions of the
cluster chemical potential $\protect\mu ^{\prime }$ for (c) half-filled
case, $\protect\mu _{\mathrm{lat}}=4t$, and (d) away from half-filling, $%
\protect\mu _{\mathrm{lat}}=3t$ (corresponds to 0.4\% electron doping). }
\label{Fig.new_Omega}
\end{figure}

The proposed functionals Eqs.(\ref{Omega1}) and (\ref{Omega2}) have been
implemented into the VCA scheme using the Lanczos algorithm of exact
diagonalization (ED) to solve the cluster problem. The Hubbard model on a
square lattice with nearest neighbor hopping $t$ has been studied, and a
cluster of $2\times 2$ sites was used in the ED scheme. While the hopping
parameter $t$ and the chemical potential $\mu $ of the lattice model remain
fixed, we have the freedom of varying the corresponding parameters of the
cluster $t^{\prime }$ and $\mu ^{\prime }$, whose optimal values should be
found by solving the stationarity equation (\ref{d_cluster}).

Figure~\ref{Fig.new_Omega}a shows the comparison between the original grand
potential $\Omega $ (dotted line) and the two potentials derived in Section~%
\ref{Sec.implement}, as functions of the variational cluster hopping
parameter $t^{\prime }$. All three functionals appear to have minimum at the
same value of $t^{\prime }=t=1$ and the proposed new functionals are convex
functions of $t^{\prime }$ near the solution, as is the original grand
potential $\Omega $.

The dependence of the grand potential on the cluster chemical potential $\mu
^{\prime }$ is plotted in Fig.~\ref{Fig.new_Omega}b for the half-filled
case. Functional $\Omega _{1}$ shown by the solid line, clearly is a convex
function near the stationary solution $\mu _{0}^{\prime }=4t$, unlike the
original grand potential $\Omega $ (dotted line) that develops a maximum at
this point. The details of the behavior of the new functionals near $\mu
_{0}^{\prime }$ appear more clearly in the blow up shown in Fig.~\ref%
{Fig.new_Omega}c. The functional $\Omega _{2}$ (dashed line) has vanishing
second derivative within machine precision around the stationary solution.
The same conclusion equally holds away from half-filling, as illustrated in
Fig.~\ref{Fig.new_Omega}d.

The above illustrations as well as many tests performed for different values
of the parameters $U$, $t$ and $\mu $ all show that the proposed functional $%
\Omega _{1}$ given by Eq.~(\ref{Omega1}) develops a minimum at the
stationary solution with respect to both variational parameters $\mu
^{\prime }$ and $t^{\prime }$. The functional $\Omega _{2}$, is less useful
for practical calculations since it exhibits a vanishing second derivative
with respect to $\mu ^{\prime }$ sufficiently close to half-filling.

Despite these encouraging results, note that the actual values of $\Omega
_{1}$ and $\Omega _{2}$ at the solution, while being nearly equal, deviate
appreciably from the original grand potential $\Omega $. There would be no
such deviation if we could have evaluated $\mathrm{f}[\Sigma ]$ exactly
instead of approximately as we were forced to do.

\section{Connection to the Chitra--Kotliar functional}

\label{Sec.CK}

The purpose of this Appendix is to identify the connection that exists
between Potthoff's original functional~\cite{Potthoff03}, the new convex
functional $\Omega \lbrack \Sigma ]$ proposed in this work (Sec.~\ref%
{Sec.new}), and an earlier attempt to construct a convex functional
undertaken by Chitra and Kotliar in Ref.~\cite{CK01}. According to the
latter study, one can construct an improved version of the Baym-Kadanoff
functional $\Omega _{BK}[G]$ given by Eq.~(\ref{BK}) as follows: 
\begin{equation}
\Omega _{\text{CK}}[G]=\Omega _{BK}[G]-\mathrm{Tr}\ln (1+JG)+\mathrm{Tr}(JG),
\label{CK}
\end{equation}%
where $J$ is the external source field coupled to the electron's Green
function 
\begin{equation}
-J=\frac{\delta \Omega _{BK}}{\delta G}=G^{-1}-G_{0}^{-1}+\frac{\delta \Phi 
}{\delta G},
\end{equation}%
where $\Phi \lbrack G]$ is the Luttinger-Ward functional. Since $J=0$ when
the Dyson equation is satisfied, both functionals are equal to the grand
potential at the stationary solution. Chitra and Kotliar have shown~\cite%
{CK01} that their new functional has a different stability criterion for its
stationary solution from that of the Baym-Kadanoff functional, however they
have been unable to prove its convexity at the stationary point. In fact,
they have shown explicitly that, in the Hartree approximation, their
proposed functional $\Omega _{\text{CK}}$ is \emph{unstable} for repulsive
interactions. This prohibits the use of the Chitra--Kotliar functional as a
variational free energy and leaves open the question of convexity.

Is there a way to relate the Chitra--Kotliar functional Eq.~(\ref{CK}) to
the functionals of self-energy $\Omega _{1}[\Sigma ]$ and $\Omega
_{2}[\Sigma ]$ proposed in Section~\ref{Sec.new}? From the previous two
equations, it is easy to show that (see Eq.~(12) in Ref.~\cite{CK01}): 
\begin{equation}
\Omega _{\text{CK}}[G]=\Phi \lbrack G]-\mathrm{Tr}\left( G\frac{\delta \Phi 
}{\delta G}\right) -\mathrm{Tr}\ln \left[ G_{0}^{-1}-\frac{\delta \Phi }{%
\delta G}\right] .  \label{CK2}
\end{equation}%
Using the relation (\ref{LW}), $\delta \Phi /\delta G=\Sigma \lbrack G]$,
and the Legendre transform of the Luttinger-Ward functional \cite{Potthoff03}
as in Eq.~(\ref{FSigma}) 
\begin{equation}
F[\Sigma ]\equiv \Phi \left[ G[\Sigma ]\right] -\mathrm{Tr}\left( \Sigma
G\right) ,
\end{equation}%
$\Omega _{\text{CK}}$ in Eq.~(\ref{CK2}) becomes 
\begin{equation}
\Omega _{\text{CK}}[G(\Sigma )]=F[\Sigma ]-\mathrm{Tr}\ln \left[
G_{0}^{-1}-\Sigma \right]  \label{CK3}
\end{equation}%
This last expression is nothing else but Potthoff's functional $\Omega
\lbrack \Sigma ]$ as given in Eq.~(\ref{GPSigma}). In other words, we see
that the Chitra--Kotliar functional of $G$, if expressed in terms of the
self-energy $\Sigma $, is equivalent to Potthoff's $\Omega \lbrack \Sigma ]$%
. Although both groups begin with the Baym-Kadanoff functional to propose
their own functional, they use the Dyson equation in apparently different
but in fact equivalent ways.

As regards the new functionals of the self-energy introduced in this work
(see Section~\ref{Sec.new}), their derivation, despite being similar in
spirit to that of Chitra and Kotliar~\cite{CK01}, have required additional
ingredients that are contained neither in the Potthoff's grand potential $%
\Omega \lbrack \Sigma ]$, nor in the Chitra--Kotliar functional.

\bibliographystyle{apsrev}
\bibliography{citations}

\end{document}